\documentclass[twocolumn, fontsize=9]{scrartcl}

\usepackage{mathtools}
\usepackage[colorlinks = true, linkcolor=black, citecolor=black, urlcolor=black]{hyperref}
\usepackage{authblk}
\usepackage[title]{appendix}
\usepackage{caption}
\usepackage[
    backend=biber,
    style=nature,
    natbib=true,
    sortlocale=en\_US,
    url=false, 
    doi=true,
    eprint=false
]{biblatex}
\AtEveryBibitem{\clearfield{issn}}
\AtEveryCitekey{\clearfield{issn}}
\AtEveryBibitem{\clearfield{month}}
\AtEveryCitekey{\clearfield{month}}
\AtEveryBibitem{\clearfield{day}}
\AtEveryCitekey{\clearfield{day}}
\AtEveryBibitem{\clearfield{issue}}
\AtEveryCitekey{\clearfield{issue}}
\begin{document}

\title{Precision spectroscopy on $^9\text{Be}$ overcomes limitations from nuclear structure}

\author[1,*]{Stefan Dickopf}
\author[1]{Bastian Sikora}
\author[1]{Annabelle Kaiser}
\author[1]{Marius Müller}
\author[2,3]{Stefan Ulmer}
\author[1]{Vladimir A. Yerokhin}
\author[1]{Zoltán Harman}
\author[1]{Christoph H. Keitel}
\author[1]{Andreas Mooser}
\author[1]{Klaus Blaum}
\date{}

\affil[1]{Max Planck Institute for Nuclear Physics, Saupfercheckweg 1, Heidelberg, 69117, Germany}

\affil[2]{Institute for Experimental Physics, Heinrich Heine University Düsseldorf, Düsseldorf, 40225, Germany}

\affil[3]{Ulmer Fundamental Symmetries Laboratory, RIKEN, Saitama, 351-0198, Japan}

\affil[*]{\href{mailto:stefan.dickopf@mpi-hd.mpg.de}{stefan.dickopf@mpi-hd.mpg.de}}

\makeatletter
\twocolumn[
   \begin{@twocolumnfalse} 
     \maketitle 
     \begin{abstract}
    Many powerful tests of the Standard Model of particle physics and searches for new physics with precision atomic spectroscopy are plagued by our lack of knowledge of nuclear properties.
    Ideally, such properties may be derived from precise measurements of the most sensitive and theoretically best-understood observables, often found in hydrogen-like systems.
    While these measurements are abundant for the electric properties of nuclei, they are scarce for the magnetic properties, and precise experimental results are limited to the lightest of nuclei~\cite{Hellwig1970, Wineland1972, Mathur1967, Schneider2022}.
    Here, we focus on $^9\text{Be}$ which offers the unique possibility to utilize comparisons between different charge states available for high-precision spectroscopy in Penning traps to test theoretical calculations typically obscured by nuclear structure.
    In particular, we perform the first high-precision spectroscopy of the $1s$ hyperfine and Zeeman structure in hydrogen-like $^9\text{Be}^{3+}$.
    We determine its effective Zemach radius with an uncertainty of $500$ ppm,
    and its bare nuclear magnetic moment with an uncertainty of $0.6$ parts-per-billion (ppb) - uncertainties unmatched beyond hydrogen.
    Moreover, we compare to measurements conducted on the three-electron charge state $^9\text{Be}^{+}$~\cite{Wineland1983}, which, for the first time, enables testing the calculation of multi-electron diamagnetic shielding effects of the nuclear magnetic moment at the ppb level.
    In addition, we test quantum electrodynamics (QED) methods used for the calculation of the hyperfine splitting.
    Our results serve as a crucial benchmark essential for transferring high-precision results of nuclear magnetic properties across different electronic configurations.
    \vspace{\baselineskip}
     \end{abstract}
    \end{@twocolumnfalse}
]
\makeatother
\section*{}\label{sec1}
\vspace{-1\baselineskip + -8.5pt}
Historically, advances in precision atomic spectroscopy have progressed alongside the development of the quantum theories of nature to successfully explain even the smallest of contributions to atomic transition frequencies.
Meanwhile, precise tests and searches for new physics require that the established theories provide sufficiently accurate predictions for the observed system.
Nowadays, in atomic systems, this frequently necessitates the knowledge of nuclear properties.
However, such properties are often not known accurately enough from theoretical models describing nuclear structure and are instead determined experimentally from independent measurements with high sensitivity to nuclear structure.
For example, measurements of the Lamb shift in atomic~\cite{Bezginov2019} or muonic hydrogen~\cite{Antognini2013} and deuterium~\cite{Pohl2016} are 
combined with other transitions to independently determine the proton charge radius and the Rydberg constant~\cite{Beyer2017} - however, discrepancies between competing results remain unsolved~\cite{Gao2022}.
Likewise, the magnetic dipole-dipole interaction of the nucleus with the bound electrons, which results in the hyperfine splitting (HFS), depends on the magnetic moment of the nucleus and is significantly influenced by the Zemach radius - a measure of electric and magnetic form factors of the nucleus.
Here, measurements of the $1s$-HFS interval in hydrogen-like systems are the most sensitive to the Zemach radius and serve as ideal references to evaluate the nuclear structure effects in other HFS intervals and test QED~\cite{Karshenboim2005}.
However, for low nuclear charge $Z$, such measurements exist only for the hydrogen isotopes~\cite{Karshenboim2005} and $^3\text{He}$~\cite{Schneider2022, Patkos2023}, while for high $Z$, tests of the HFS lack accurate experimental values of the nuclear magnetic moments~\cite{Skripnikov2018, Shabaev2000_Bi, Stone2019}.

Recently, the high-precision Penning-trap measurement of the Zeeman and hyperfine splitting of $^3\text{He}^{+}$ allowed to directly measure its Zemach radius and nuclear magnetic moment~\cite{Schneider2022}, simultaneously providing both parameters needed for precise predictions of other HFS intervals~\cite{Patkos2023}.
Additionally, the accurate value of the magnetic moment of the atom enables absolute magnetometry with hyperpolarized $^3\text{He}$, ref.~\cite{Farooq2020}.
However, this requires transferring the measured nuclear magnetic moment from the hydrogen-like system to the neutral system, which involves the theoretical calculation of diamagnetic shielding parameters.
In the past, inadequate calculations of such parameters have led to several discrepancies in precision physics~\cite{Crespo1996, Fella2020, Ullmann2017, Skripnikov2018}, including the recent $7$-$\sigma$ deviation of the HFS specific difference in $^{209}\text{Bi}^{82+,80+}$.
In these studies, the required nuclear magnetic moments were obtained from measurements using systems with complex electronic structure and calculations of shielding parameters relying on quantum chemistry codes which frequently provide no, or underestimated, uncertainties~\cite{Stone2019}.
In contrast, for systems such as hydrogen-like and neutral $^3\text{He}$, these issues are remedied by the simple electronic structure, which enables diamagnetic shielding calculations using highly accurate nonrelativistic quantum electrodynamics (NRQED) methods.
Here, the diamagnetic shielding parameters are calculated in a perturbative approach and the estimated uncertainties are more than one order of magnitude better than the experimental value of the $^3\text{He}$ nuclear magnetic moment~\cite{Wehrli2021}.
However, adjustments to the NRQED theory value at the same level as the experimental uncertainty were performed recently~\cite{Pachucki2023_3}, further motivating an experimental verification and benchmark for diamagnetic shielding calculations.

An ideal candidate to both test the diamagnetic shielding calculations and introduce a highly accurate reference for nuclear structure contributions in the hyperfine interaction is $^9\text{Be}$.
Here, the low nuclear charge of $Z=4$ permits calculations of the highest available accuracy in the hydrogen-like system while, simultaneously, the Zeeman and hyperfine splitting can be probed via high-precision spectroscopy in Penning traps for both the lithium-like, $^9\text{Be}^{+}$, and hydrogen-like, $^9\text{Be}^{3+}$, charge states.
In this work, we present the first measurement on $^9\text{Be}^{3+}$.
Compared to extractions using $^9\text{Be}^+$~\cite{Wineland1983, Pachucki2010}, the higher accuracy of theoretical calculations possible in hydrogen-like $^9\text{Be}^{3+}$ allows for significantly improved determinations of the Zemach radius as well as the magnetic moment of the bare nucleus.
Additionally, we perform a unique comparison between the experimental hyperfine and Zeeman splitting of $^9\text{Be}^{+}$ and $^9\text{Be}^{3+}$, which we use to eliminate the nuclear structure-dependent terms.
Compared to the measurements of the hyperfine splittings of $^{209}\text{Bi}^{82+,80+}$, ref.~\cite{Skripnikov2018}, this not only allows testing the QED theory via the HFS specific difference, but also testing calculations of multi-electron diamagnetic shielding parameters at the ppb level.
The latter constitutes the first precision test of the corrections to nuclear magnetic moments across different charge states.

\begin{figure}[ht!]
    \includegraphics{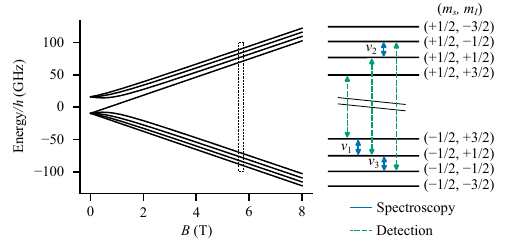}
    \caption{
    \textbf{Hyperfine and Zeeman splitting of $^9\text{Be}^{3+}$.}
    On the left, the energies of the spin states are shown as a function of the external magnetic field $B$.
    The level structure at our magnetic field (dashed box) of around $5.7~\text{T}$ is shown on the right. 
    Spectroscopy is performed on the three nuclear spin transitions (blue) labeled $\nu_1$, $\nu_2$ and $\nu_3$.
    The three electron spin transitions (green) are used for detecting the nuclear spin state.
    \label{fig:breit_rabi}}
\end{figure}

The combined hyperfine and Zeeman interaction in $^9\text{Be}^{3+}$ is described by the Hamiltonian
\begin{equation}
    H = - \frac{1}{2\pi}\frac{e}{2 m_e} g_s B S_z - \frac{1}{2\pi}\frac{e}{2 m_p} g_I^{\prime} B I_z + \nu_{\text{HFS}} \boldsymbol{S}\cdot\boldsymbol{I},
    \label{eq:breit_rabi}
\end{equation}
where $e$ is the elementary charge and $m_e$, $m_p$ are the electron and proton mass, respectively, and $\nu_{\text{HFS}}$ is the hyperfine splitting.
The external magnetic field $B$ is chosen to define the $z$-direction as the quantization axis for the spin angular momenta of the electron $\boldsymbol{S}$ and nucleus $\boldsymbol{I}$.
In this formula, the magnetic moments of the bound electron and shielded nucleus are given in units of the Bohr and nuclear magneton via the gyromagnetic ratios ($g$-factors), $g_s$ and $g^{\prime}_I$ (the prime indicates the shielding), respectively.
Using the spin magnetic quantum numbers of the electron, $m_s$, and the nucleus, $m_I$, the level structure of $^9\text{Be}^{3+}$ is visualized in Figure~\ref{fig:breit_rabi}.
The twelve magnetic dipole transitions can be split into six low-frequency, $5.7~\text{GHz}< \nu < 7.2~\text{GHz}$, nuclear spin transitions with $(\Delta m_s, \Delta m_I) = (0, 1)$, four high-frequency, $141~\text{GHz}< \nu < 181~\text{GHz}$, electron spin transitions with $(\Delta m_s, \Delta m_I) = (1, 0)$, and two high-frequency combined transitions with $(\Delta m_s, \Delta m_I) = (1, 2)$.
The highest sensitivity for the extraction of $g^{\prime}_I$ and $\nu_{\text{HFS}}$ is reached with the measurement of the two nuclear transitions $\nu_1 \approx 6.622~\text{GHz}$ and $\nu_2\approx 6.553~\text{GHz}$, compare Figure~\ref{fig:breit_rabi}.
Our determination of the magnetic field $B$ requires the precise knowledge of the mass $m_{^9\text{Be}^{3+}}$ of the $^9\text{Be}^{3+}$ ion.
Since the uncertainty $\delta m_{^9\text{Be}}/m_{^9\text{Be}} = 9\times10^{-9}$ of the current accepted mass value~\cite{AME2020} would limit the extraction of $g_I^{\prime}$, the measurement of a third nuclear transition $\nu_3 \approx 6.124~\text{GHz}$ was included, allowing us to independently determine the mass.

\begin{figure*}[ht!]
    \includegraphics[width = \linewidth]{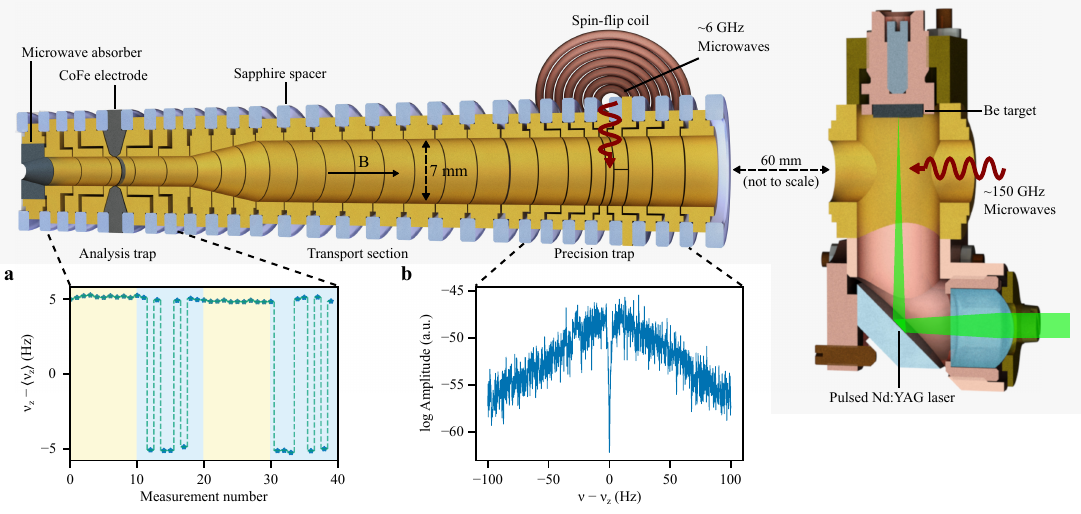}
    \caption{
    \textbf{Schematic of the Penning-trap setup.}
    The Penning trap is built up of gold-plated copper electrodes separated by isolating sapphire spacers (blue).
    Microwaves for driving the electron spin transitions are irradiated on-axis via a waveguide and a coil with a few windings connected to coaxial cables is used to drive the low-frequency nuclear transitions in the precision trap.
    $^9\text{Be}^{3+}$ ions are produced via laser ablation from a solid beryllium target and subsequent electron impact ionization.
    \textbf{a}, While alternating the irradiation of the two electron spin detection transitions (indicated by the two background colors) in the analysis trap, axial frequency jumps are observed only for one of them.
    \textbf{b}, The motional frequencies of $^9\text{Be}^{3+}$ are measured via `dip` signals in the Fourier spectrum of the detection signal.
    \label{fig:setup}}
\end{figure*}
The measurements are performed with a single ion in the Penning-trap setup shown in Figure~\ref{fig:setup}.
The setup is placed in a sealed-off vacuum chamber inside the liquid helium-cooled bore of a $5.7$-T superconducting magnet.
Inside this chamber, the vacuum conditions allow for trapping lifetimes exceeding the multiple months needed for the full measurement campaign.
The magnetic field confines the ion on a circular orbit with revolution frequency $\nu_c = (q B)/(2\pi m_{^9\text{Be}^{3+}})$, where $\nu_c$ is the cyclotron frequency and $q=3\, e$ is the charge of $^9\text{Be}^{3+}$.
The cylindrical electrodes of the trap are biased by an ultrastable voltage source to create a quadrupolar electrostatic potential, forcing the ion into a harmonic oscillation along the $z$-axis with frequency $\nu_z$.
This further splits the radial motion into two eigenmotions characterized by the modified cyclotron frequency $\nu_+$ and the magnetron frequency $\nu_-$.
Here $\nu_c \approx \nu_+ \approx 29~\text{MHz} \gg \nu_z \approx 480~\text{kHz} \gg \nu_- \approx 4~\text{kHz}$. 
By measuring the three eigenfrequencies and combining them via the so-called invariance theorem, $\nu_c^2 = \nu_+^2 + \nu_-^2 + \nu_z^2$, the cyclotron frequency is reproduced while canceling certain systematic effects~\cite{GABRIELSE2009}.
A superconducting tank circuit is connected to one of the trap electrodes to provide resistive cooling of the axial mode to the ambient $4.2~\text{K}$ as well as detection of the axial oscillation, see Figure~\ref{fig:setup} and ref.~\cite{WINELAND1975}.
Sideband coupling to the axial mode enables thermalization and frequency measurement of the radial modes~\cite{Cornell1990}.

For the spectroscopy of the spin transitions, we count spin-state changes of the ion following an excitation with a frequency close to the transition center.
Changes of the spin state are detected by employing the continuous Stern-Gerlach effect~\cite{Dehmelt1986}.
To this end, a ferromagnetic ring electrode introduces a quadratic magnetic field $\Delta B = B_2 z^2$, where $B_2 \approx 282~\text{kTm}^{-2}$.
This couples the ion's magnetic moment to its axial motion, slightly altering $\nu_z$ depending on the spin state.
Changes of the spin state induced by a transition $(m_s, m_I) \rightarrow (m_s^{\prime}, m_I^{\prime})$ lead to a shift $\Delta\nu_z$ proportional to the change of the magnetic moment~\cite{Dehmelt1986}.
For an electron spin transition, $(\Delta m_s, \Delta m_I) = (1, 0)$, the axial frequency jump $\Delta \nu_z \approx 10~\text{Hz}$ can be easily detected.
In contrast, the change of the ion's magnetic moment for a nuclear spin transition,  $(\Delta m_s, \Delta m_I) = (0, 1)$, is greatly reduced, rendering its detection challenging.
For instance, in the case of $\nu_1$, the axial frequency jump is only $\Delta \nu_z \approx 6~\text{mHz}$, which cannot be discerned from the background fluctuations of $\nu_z$.
Instead, for the spectroscopy of nuclear transitions $(m_s, m_I) \rightarrow (m_s, m_I^{\prime})$, the two detection transitions $(m_s, m_I) \rightarrow (m_s^{\prime}, m_I)$ and $(m_s, m_I^{\prime}) \rightarrow (m_s^{\prime}, m_I^{\prime})$ are used, compare Figure~\ref{fig:breit_rabi}.
While cycling these two transitions only one of them produces detectable electron spin-state changes, see Figure~\ref{fig:setup}\,\textbf{a}, which unambiguously identifies the nuclear spin state.

Since the large $B_2$ required for spin-state detection would limit the experimental precision, we utilize spatially separated traps for spin-state detection and the precision measurement, called analysis trap (AT) and precision trap (PT)~\cite{Haeffner2000}, see Figure~\ref{fig:setup}.
In the PT, the residual magnetic field inhomogeneity is greatly reduced, $B_{2, \text{PT}} \approx 1~\text{T}/\text{m}^2$.
A measurement cycle starts by determining the spin state in the AT.
Following an adiabatic transport to the PT, an initial cyclotron frequency measurement $\nu_{c,1}$ determines the expected spin transition frequency.
During a second measurement of the cyclotron frequency $\nu_{c,2}$, the spin transition is driven with a frequency $\nu_{\text{MW}}$, which is randomly offset from the previously calculated value.
After a third measurement, $\nu_{c,3}$, the ion is transported back to the AT to again detect the spin state and determine whether it changed from the previously detected one.
We measured $\nu_c$ to a precision of one part in a billion with typical averaging times of a few minutes and performed a single measurement cycle in 20 minutes.
For each of the three transitions, a few hundred measurement cycles were performed.

\begin{figure*}[ht!]
\centering
    \includegraphics{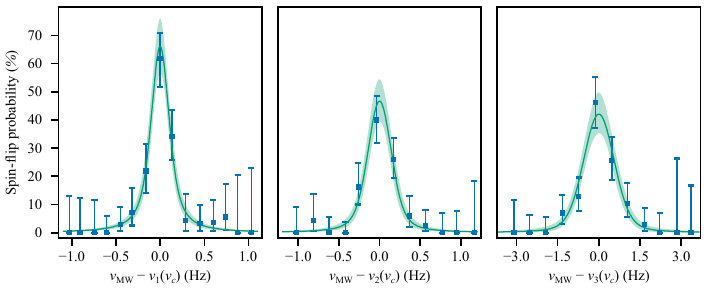}
    \caption{\textbf{Recorded resonance curves.}
       The difference of the probe excitation at frequency $\nu_{\text{MW}}$ to the nuclear spin transition frequency $\nu_i(\nu_c$) is calculated from the simultaneously measured $\nu_c$.
       The data points are the binned number of successful spin-flip tries divided by the total number of tries in that bin, and the error bars correspond to the $68\%$-binomial confidences.
       The line and confidence band ($68\%$) are taken from a Voigt profile fit via maximum likelihood analysis of the unbinned data.
       \label{fig:resonances}}
\end{figure*}
We use maximum likelihood estimation to fit the center values of $\nu_{\text{MW}} - \nu_i(\nu_c \vert \Gamma_e, \Gamma_I, \nu_{\text{HFS}})$ from the three recorded resonances shown in Figure~\ref{fig:resonances} to extract $\Gamma_e$, $\Gamma_I$ and $\nu_{\text{HFS}}$, where
\begin{equation}
    \label{eq:fit_parameters}
    \Gamma_e = \frac{g_s}{2}\frac{e}{q}\frac{m}{m_e},\;\; \Gamma_I = \frac{g_I^{\prime}}{g_s}\frac{m_e}{m_p},
\end{equation}
compare Eq.~(\ref{eq:breit_rabi}) and 
see Supplementary Information for details.
The statistical uncertainties are $17~\text{mHz}$, $26~\text{mHz}$ and $86~\text{mHz}$ for $\nu_1$, $\nu_2$ and $\nu_3$, respectively.
Experimental systematic shifts and uncertainties are due to special relativity effects, electrostatic and magnetostatic field imperfections, induced image charges, the axial frequency determination, and the accuracy of the GPS-locked rubidium clock.
Additionally, we include a second-order effect arising from the electric quadrupole moment of the nucleus~\cite{Moskovkin2006}, slightly shifting the transition frequencies.
A discussion of systematic effects and the error budget is found in the Supplementary Information.
The corrected results are $\Gamma_e = -5479.8633435(11)(19)$, $\Gamma_I = 2.1354753854(11)(3)\times 10^{-4}$ and $\nu_{\text{HFS}} = -12796971342.630(50)(15)$~Hz, where the first number in parentheses is the statistical and the second the systematic uncertainty.

Both the $g$-factor of the free electron and of the bare nucleus need to be corrected for the influence of the binding potential in the composite system~\cite{Faustov1970}.
Our calculations of the bound-electron $g$-factor include corrections due to special relativity, QED, nuclear recoil, and structure effects (see Supplementary Information).
We evaluate $g_s(^9\text{Be}^{3+})= -2.0017515747$, where uncertainties due to uncalculated higher-order QED corrections as well as nuclear corrections are smaller than 1 in the last given digit.
The binding corrections to the bare nuclear $g$-factor $g_I$ are expressed as $g_I^{\prime} = (1-\sigma)g_I$, where $\sigma$ is the diamagnetic shielding parameter.
The theoretical calculations of diamagnetic shielding parameters include corrections due to nuclear recoil, relativistic effects, one-loop QED, and finite nuclear size effects.
We evaluate $\sigma(^9\text{Be}^{3+}) = 71.15397(14)\times 10^{-6}$ (see the Supplementary Information).

From $\Gamma_I$, we calculate the bare nuclear $g$-factor using the proton-to-electron mass ratio~\cite{Codata2018}, our values of the bound-electron $g$-factor $g_s(^9\text{Be}^{3+})$ and the shielding $\sigma(^9\text{Be}^{3+})$.
Our result, $g_I = -0.78495442296(42)_{\text{exp}}(11)_{\text{theo}}$, improves the accuracy by a factor of $45$ compared to the result from ref.~\cite{Pachucki2010}, the latter being limited by the shielding parameter of $^9\text{Be}^{+}$.
With a fractional uncertainty of $0.6~\text{ppb}$, our result establishes the nuclear magnetic moment of $^9\text{Be}$ as the second most precise, surpassed only by that of the proton~\cite{Schneider2017}.
Via comparisons to $^9\text{Be}^{+}$ we evaluate the shielding factor of the lithium-like system,
\begin{equation}
    1 - \sigma(^9\text{Be}^{+}) = (1-\sigma(^9\text{Be}^{3+}))\frac{\Gamma_I(^9\text{Be}^{+})}{\Gamma_I(^9\text{Be}^{3+})}\frac{g_s(^9\text{Be}^{+})}{g_s(^9\text{Be}^{3+})}.
\end{equation}
This requires the experimental result of $\Gamma_I(^9\text{Be}^{+})$ from ref.~\cite{Wineland1983} and the bound-electron $g$-factor $g_s(^9\text{Be}^{+})=-2.0022621287(24)$.
For the latter, we use the calculations performed in ref.~\cite{Pachucki2010} and the updated nuclear recoil correction~\cite{Pachucki2023, Shabaev2017}.
We evaluate $\sigma(^9\text{Be}^{+}) = 141.8821(11)_{\text{exp}}(12)_{\text{theo}}\times 10^{-6}$, where the second uncertainty is limited by $g_s(^9\text{Be}^{+})$.
The theoretical value, $\sigma(^9\text{Be}^{+})_{\text{theo}} = 141.85(3)\times 10^{-6}$, ref.~\cite{Pachucki2010}, is in good agreement with our experimental result.
To our knowledge, this constitutes the first high-precision test of the calculation of a multi-electron diamagnetic shielding parameter.
The shielding calculations for the three-electron systems $^9\text{Be}^{+}$ and $^{6,7}\text{Li}$ are performed identically and use explicit values of the leading- and lowest-order recoil terms and an estimate of the relativistic correction~\cite{Pachucki2023, Pachucki2010}.
At the current state of theoretical calculations of the lithium-like shielding parameters, we confirm the leading-order calculation and the estimate of the relativistic correction, solidifying its use for $^{6,7}\text{Li}$.
In the future, the advanced calculations performed for $^3\text{He}$ can be extended to $^9\text{Be}^{+}$, ref.~\cite{Wehrli2022}, and our experimental value of $\sigma(^9\text{Be}^{+})$ will serve as an ideal benchmark at the ppb precision level.

Calculations of the zero-field hyperfine splitting can be expressed as~\cite{Puchalski2014, Sun2023, Pachucki2023_2}
\begin{equation}
    \nu_{\text{HFS}} = \frac{E_{\text{F}}}{2 h} \left(1 + \delta_{\text{pt}} - 2 Z \tilde{r}_Z/a_0\right),
\end{equation}
where $E_{\text{F}}$ is the non-relativistic value of the hyperfine splitting~\cite{Fermi1930}, $a_0$ is the Bohr radius, $\delta_{\text{pt}}$ summarizes all corrections with a point-like treatment of the nucleus (see Supplementary Information), and all nuclear structure contributions are absorbed in $-2Z\tilde{r}_Z/a_0 \approx 6 \times 10^{-4}$ via the effective Zemach radius $\tilde{r}_Z$.
In comparison, the relative nuclear structure contributions to the electron $g$-factor $g_s$ and the shielding $\sigma(^9\text{Be}^{3+})$ are below $10^{-10}$, which highlights the sensitivity of the HFS to effects from nuclear structure.
From our experimental result, $\nu_{\text{HFS}}(^9\text{Be}^{3+}) = -12796.971342630(52)~\text{MHz}$, we calculate the effective Zemach radius $\tilde{r}_Z = 4.048(2)~\text{fm}$.
This value is consistent with the one extracted from $^9\text{Be}^{+}$, $\tilde{r}_Z = 4.03(5)~\text{fm}$~\cite{Puchalski2014} (value corrected with our more accurate magnetic moment), and improves its accuracy by a factor of $25$, which is possible due to the more accurate calculation of $\delta_{\text{pt}}(^9\text{Be}^{3+})$ compared to $\delta_{\text{pt}}(^9\text{Be}^{+})$.

Following investigations on $^{209}\text{Bi}$, ref.~\cite{Ullmann2017}, we form a specific difference between the hydrogen- and lithium-like systems, $\Delta\nu_{\text{HFS}} = \nu_{\text{HFS}}(^9\text{Be}^{+}) - \xi \nu_{\text{HFS}}(^9\text{Be}^{3+})$ to cancel the large theoretical uncertainties due to nuclear structure with the calculated weighting factor $\xi = 0.04881891046$ (Supplementary Information).
This complements the high-$Z$ case of $^{209}\text{Bi}$ due to the different theoretical approach used to calculate the lithium-like HFS.
For high $Z$, the large relativistic effects are included directly in the leading order via the use of a relativistic wave function basis set, while electron-electron correlations are treated perturbatively~\cite{Volotka2012}.
In contrast, for low $Z$, the electron-electron correlations lead to considerably larger contributions which requires using wave functions constructed from an explicitly correlated, non-relativistic basis set, and the relativistic corrections are instead treated perturbatively~\cite{Yerokhin2008, Puchalski2013}.
Concluding, in the case of $^{209}\text{Bi}$, higher-order QED terms are tested via $\Delta\nu_{\text{HFS}}$, while in our study at low $Z$, higher-order electron correlations effects are benchmarked instead.
The experimental result, using $\nu_{\text{HFS}}(^9\text{Be}^{+}) = -625.008837044(12)~\text{MHz}$ from ref.~\cite{Shiga2011}, is $\Delta\nu_{\text{HFS, exp}} = -274.638909(12)~\text{kHz}$, where the uncertainty is dominated by $\nu_{\text{HFS}}(^9\text{Be}^{+})$.
We calculate the theoretical value, $\Delta\nu_{\text{HFS, theo}} = -271.4(3.6)~\text{kHz}$, which is in good agreement with the experimental value, but has a much larger uncertainty.
Equivalently, this constitutes a test of $\nu_{\text{HFS}}(^9\text{Be}^{+})$ with $6$-ppm precision.
Similarly to calculations for $^{6,7}\text{Li}$, estimates of the QED contributions to the lithium-like system limit the accuracy of the theoretical result~\cite{Pachucki2023_2}.

\begin{figure}[ht!]
    \includegraphics{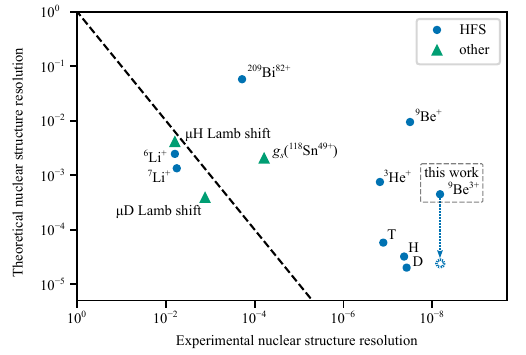}
    \caption{\textbf{Theoretical and experimental nuclear structure resolution of various measurements}~\cite{Hellwig1970, Wineland1972, Mathur1967, Schneider2022, Ullmann2017, Shiga2011, Sun2023, Guan2020, Antognini2013, Pohl2016, Morgner2023, Shabaev2000_Bi, Puchalski2014, Pachucki2023_2}. The $y$-axis is the fraction of the theoretical uncertainty of the point-nucleus calculations to the nuclear structure contribution. The $x$-axis is the fraction of experimental uncertainty to the nuclear structure contribution. The dashed diagonal line indicates equal experimental and theoretical resolution. For H, D, and T, HFS calculations from ref.~\cite{Pachucki2023_2} were used, which, opposed to our value for $\nu_{\text{HFS}}(^9\text{Be}^{3+})$, do not include contributions and uncertainties from hadronic and muonic vacuum polarization as well as certain nuclear recoil terms. For comparison, the blue, dashed line and point indicate the reduced theoretical uncertainty of $\nu_{\text{HFS}}(^9\text{Be}^{3+})$ using the same calculations which do not include the aforementioned contributions. For details, see Table~S7 in the Supplementary Information. \label{fig:results}}
\end{figure}
In Figure~\ref{fig:results}, we compare the nuclear structure resolution of several measurements.
Our determination of the nuclear structure contribution to the HFS of $^9\text{Be}$ via $^9\text{Be}^{3+}$ shows leading experimental resolution and significant improvements of the theory compared to $^9\text{Be}^{+}$.
This progress now enables testing other HFS systems in $^9\text{Be}$ with higher accuracy using our precise values of the Zemach radius and the nuclear magnetic moment, as demonstrated by our evaluation of the specific difference.
Furthermore, future tests of the theory of the HFS in the helium-like system $^9\text{Be}^{2+}$ will benefit from our results~\cite{Qi2023} and direct measurements of the $2s$ HFS in $^9\text{Be}^{3+}$ would provide an opportunity for highly stringent tests of bound-state QED due to the more accurate theory in single-electron systems.

Lastly, we calculate the atomic mass of $^9\text{Be}$ from $\Gamma_e$, using the electron mass ~\cite{Codata2018}, the required binding energies~\cite{Kramida2005, drake1988theoretical, beigang1983} and our theory value of $g_s(^9\text{Be}^{3+})$.
The result, $m_{^9\text{Be}} = 9.0121830344(35)~\text{u}$, is in perfect agreement with the accepted value~\cite{AME2020} and improves the uncertainty by a factor of $20$.

\begin{table}[!h]
  \caption{\label{tab:results}\raggedright \textbf{Summary of results.} The label `tw` refers to results derived in this work.}
    \begin{tabular}{l r}
    \hline
      \textbf{Value} & \textbf{Refs.} \\\hline
      \rule{0pt}{3ex}$g_I = -0.78495442296(42)_{\text{exp}}(11)_{\text{theo}}$ & tw\\
      $g_I = - 0.78495439(2)_{\text{theo}}$ & \cite{Pachucki2010}\\
      $\sigma(^9\text{Be}^{+}) = 141.8821(11)_{\text{exp}}(12)_{\text{theo}}\times 10^{-6}$ & tw, \cite{Wineland1983, Pachucki2010, Shabaev2017} \\
      $\sigma(^9\text{Be}^{+}) = 141.85(3)_{\text{theo}}\times 10^{-6}$ & \cite{Pachucki2010}\\ \hline
      \rule{0pt}{3ex}$\tilde{r}_Z = 4.048(2)~\text{fm}$ & tw\\
      $\tilde{r}_Z = 4.03(5)~\text{fm}$ & \cite{Puchalski2014}\\
      $\Delta\nu_{\text{HFS, exp}} = -274.638909(12)~\text{kHz}$ & tw,  \cite{Shiga2011}\\
      $\Delta\nu_{\text{HFS, theo}} = -271.4(3.6)~\text{kHz}$ & tw,  \cite{Puchalski2014}\\ \hline
      \rule{0pt}{3ex}$m_{^9\text{Be}} = 9.0121830344(35)~\text{u}$ & tw \\
      $m_{^9\text{Be}} = 9.01218306(8)~\text{u}$ & \cite{AME2020} \\
    \hline
    \end{tabular}
\end{table}
The results are summarized in Table~\ref{tab:results}.
In conclusion, our precision measurement of the bare nuclear magnetic moment and the effective Zemach radius of $^9\text{Be}$ with hydrogen-like $^9\text{Be}^{3+}$ enables tests of QED methods only accessible with the accurate knowledge of these properties.
At present, via the comparison to measurements on $^9\text{Be}^{+}$, we provide the first high-precision test of multi-electron diamagnetic shielding calculations and a $6$-ppm test of the QED calculations of the lithium-like $2s$ HFS.
High-precision Penning-trap measurements of hyperfine and Zeeman splittings, which we previously also demonstrated for $^3\text{He}^+$, are now possible for a multitude of other hydrogen-like or lithium-like ions, enabling the suppression of nuclear effects, mandatory for further applications such as e.g. spectroscopic searches for physics beyond the standard model~\cite{Debierre2020}.
Additionally, measurements on $^{6,7}\text{Li}^{2+}$ would enable direct comparisons to the recent measurements on the helium-like charge states~\cite{Sun2023, Guan2020, Pachucki2023_3}.

\printbibliography

\subsection*{Acknowledgments}
This work is part of and funded by the Max Planck Society and RIKEN.
Furthermore, this project has received funding from the European Research Council (ERC) under the European Union’s Horizon 2020 research and innovation programme under grant agreement no. 832848-FunI and we acknowledge funding and support by the International Max Planck Research School for Precision Tests of Fundamental Symmetries (IMPRS-PTFS) and by the Max Planck-RIKEN-PTB Center for Time, Constants and Fundamental Symmetries.
This work comprises parts of the Ph.D. thesis work of S.D. to be submitted to Heidelberg University, Germany.

\subsubsection*{Authors' contributions}
S.D., A.K., M.M., A.M. performed the measurements and B.S., V.A.Y., Z.H. carried out the QED calculations. 
The manuscript was written by S.D., B.S., S.U., V.A.Y., Z.H., C.H.K., A.M. and K.B. and discussed among and approved by all co-authors.

\end{document}


\section{Supplement - Experiment}

\subsection*{Resonance fitting and statistical uncertainties}
\begin{figure}[!h]
\centering
    \includegraphics{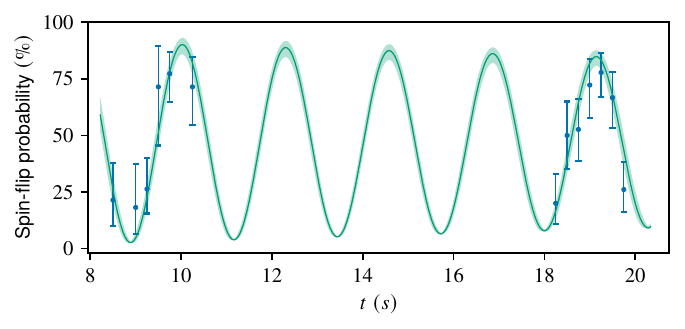}
    \caption{Measured Rabi cycle of the $\nu_1$ transition.
    The fit is done with the probability (\ref{eq:rabi_cycle}) convolved over $\Delta$ with a Gaussian.
    The shaded region shows the 1-$\sigma$ confidence band of the fit. We observe good agreement with our model.}
    \label{fig:rabi}
\end{figure}

In each measurement cycle, we measure the cyclotron frequency $\nu_c$ while irradiating microwaves at frequency $\nu_{\text{MW}}$ and subsequently determine whether a change of spin state occurred.
From the former two values we form the quantity $\tilde{\Delta}_i = \nu_{\text{MW}} - \nu_i(\nu_c, \tilde{\vec{p}})$, with a suitable guess $\tilde{\vec{p}} = (\tilde{\Gamma}_e, \tilde{\Gamma}_I, \tilde{\nu}_{\text{HFS}})$ close to the real parameters $\vec{p} = (\Gamma_e, \Gamma_I, \nu_{\text{HFS}})$.
The probability to observe a spin-state change is described by the Rabi cycle
\begin{equation}
    P(\Delta_i, t) = \frac{\Omega^2}{\Omega^2 + \Delta_i^2}\sin^2\left(\pi\sqrt{\Omega^2 + \Delta_i^2} t\right),
    \label{eq:rabi_cycle}
\end{equation}
where $\Delta_i = \nu_{\text{MW}} - \nu_i(\nu_c, \vec{p})$, $\Omega$ is the Rabi frequency and $t$ the microwave irradiation time.
The uncertainty of the magnetic field measurement through $\nu_c$ leads to a gaussian distributed $\Delta_i$ with width $\sigma(\Delta_i) \approx \frac{\partial \nu_i(\nu_c)}{\partial \nu_c} \sigma(\nu_c)$, which is added to the above probability via convolution.
In the limit of $\Omega \lessapprox \sigma(\Delta_i)$ and $t > 1/\Omega$ this leads to a Voigt profile with Gaussian width $\sigma(\Delta_i)$ and Lorentzian width $\Omega$ centered at $\Delta_i = 0$ for a fixed time $t$.
The maximum probability converges to $0.5$ for large times, but as can be seen from Fig.~3, this is not necessarily the case here.
To check our understanding of the probability lineshape we explicitly measure the Rabi cycle for the $\nu_1$ transition at a Rabi frequency $\Omega \approx 0.44~\text{Hz} > \sigma(\Delta_i) \approx 0.065~\text{Hz}$, see Figure~\ref{fig:rabi}.
With the smaller Rabi frequencies used for the resonances in Fig.~3, a weak time dependence, which was not explicitly optimized, on the reached amplitude is still expected for all transitions.

We use a maximum likelihood estimation (MLE) and a complementary Markov chain Monte Carlo method, see ref.~\cite{Foreman_Mackey_2013}, to fit the center $\langle\tilde{\Delta}_i\rangle,\; i = \{1,2,3\}$ together with the widths and amplitudes of the individual resonances with a Voigt profile.
Both methods produce identical results on the level of the significant digits of the statistical result.
To simultaneously fit the most likely values $\vec{p}$, we adjust $\tilde{\vec{p}}$ such that the fitted center values $\langle\tilde{\Delta}_i(\tilde{\vec{p}})\rangle = \langle\Delta_i(\vec{p})\rangle = 0$.

The statistical covariance matrix of the parameters is estimated via the non-linear least square estimator to be
\begin{equation}
    \label{eq:cov_stat}
    \text{cov}(\vec{p})_{\text{stat}} = J^{-1}\text{cov}(\vec{\Delta})_{\text{stat}}{J^{-1}}^{T},
\end{equation}
where 
\begin{equation}
    J_{ij}(\vec{p}) = \frac{\partial \nu_i}{\partial p_j}(\vec{p})
    \label{eq:Jacobian}
\end{equation}
is the Jacobi matrix and $\text{cov}(\vec{\Delta})_{\text{stat}} = \text{diag}(\sigma^2(\Delta_i))$ is the diagonal matrix of the variances of the fitted centers.

\subsection*{Systematic shifts}
We calculate the Zeeman and hyperfine transition frequencies via the measured cyclotron frequency $\nu_c$.
As the transition frequencies depend on the magnetic field at the position of the ion $B_{\text{ion}}$, we need to consider systematic shifts of the form $\nu_c = \frac{1}{2\pi}\frac{q}{m}B_{\text{ion}} + \delta \nu_c$.
The individual shifts of the motional frequencies are propagated by the invariance theorem $\nu_c^2 = \nu_+^2 + \nu_-^2 + \nu_z^2$ to the free cyclotron frequency.
Shifts of the free cyclotron frequency $\delta \nu_c$ then shift our fitted center by
\begin{equation}
    \langle\Delta_i\rangle \rightarrow \langle\Delta_i\rangle - \frac{\partial \nu_i}{\partial \nu_c}\delta \nu_{c} = \langle\Delta_i\rangle + \delta \nu_i.
\end{equation}
Similar to the treatment of the statistical fitting we use the Jacobian matrix, Eq.~(\ref{eq:Jacobian}), to compute the shift on the parameters as
\begin{equation}
    \vec{\delta p} = J^{-1} \vec{\delta \nu}.
\end{equation}
For the uncertainties we propagate the full covariances to the parameters, similar to the uncertainty of the statistical result, Eq.~(\ref{eq:cov_stat}),
\begin{equation}
    \text{cov}(\vec{p})_{\text{syst}} = J^{-1}\text{cov}(\vec{\delta\nu})_{\text{syst}}{J^{-1}}^{T}.
\end{equation}
Here $\text{cov}(\vec{\delta\nu})_{\text{syst}}$ is the covariance matrix of the systematic shifts on the resonance centers.
This is necessary, as we have correlated uncertainties, which include the image charge shift and uncertainties of the inhomogeneites and anharmonicity values, but also uncorrelated uncertainties from shifts depending on the thermal radii of the ion and the uncertainty of the time standard.
In particular, the shifts of $\nu_c$ lead to full cancellation in the value of $\nu_{\text{HFS}}$ and $\Gamma_I$, see Table~\ref{tab:systematics}.
Only the uncorrelated uncertainties of $\nu_c$ translate to uncertainties in $\Gamma_I$ and $\nu_{\text{HFS}}$.

\begin{figure}[!h]
\centering
    \includegraphics{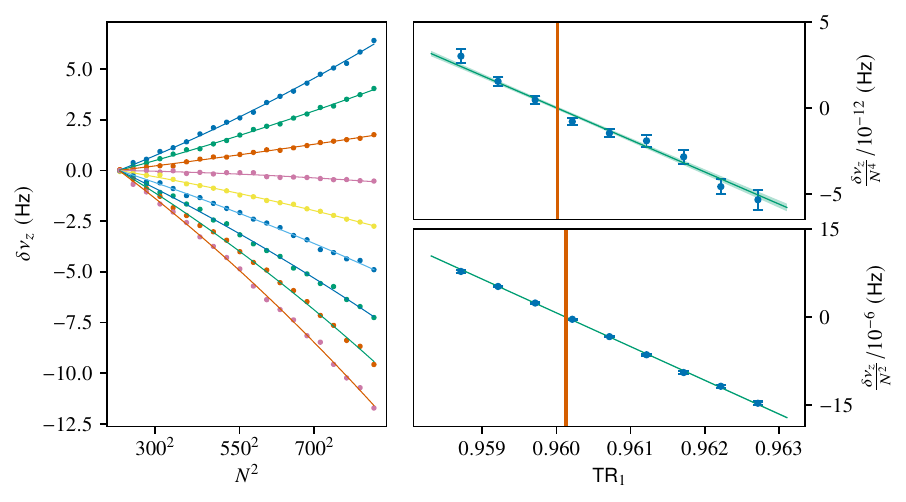}
    \caption{Optimization of the correction electrode voltages. Here, the tuning ratios $\text{TR}_{1,2}$ are defined as the ratio between the voltage on the correction electrode pair to the voltage on the central (ring) electrode. On the left, $\delta(\nu_z)(N^2)$, where $\rho_- \propto N$, is measured for different values of $\text{TR}_{1}$ and a polynomial fit is used to determine the dependencies on $\rho_-^2$ and $\rho_-^4$. On the right, these dependencies are plotted against $\text{TR}_{1}$ to determine the intercepts (colored in red), where no shifts occur. The intercept of $\delta(\nu_z)/ N^2 = 0$ is used as the optimum to give $C_4 = 0$. The small difference to the $\delta(\nu_z)/ N^4 = 0$ intercept results in a negligible residual higher order anharmonicity $C_6$.}
    \label{fig:tr_opt}
\end{figure}

For shifts due to magnetic field imperfections, only the shift of $\nu_z$ by $B_2$ due to the finite amplitude of the modified cyclotron, $\rho_+$, and magnetron mode, $\rho_-$,
\begin{equation}
    \frac{\delta \nu_z}{\nu_z} = \frac{B_2}{4 B_0}\left(\frac{\nu_c}{\nu_-}\rho_+^2 + \frac{\nu_c}{\nu_p}\rho_-^2\right),
    \label{eq:B2nuz}
\end{equation}
is relevant~\cite{Ketter2014}.
Shifts of the motional frequencies due to non-harmonic contributions in the electrostatic potential $\Phi(z) = C_2 z^2 + \sum_{k > 2} C_k z^k$ need to be considered as well.
The lowest order shifts are
\begin{align}
\begin{split}
    \delta \nu_+ &= \frac{3}{2}\frac{C_4}{C_2}\nu_-\left(-2\rho_z^2 + \rho_+^2 + 2\rho_-^2\right),\\
    \delta \nu_z &= \frac{3}{4}\frac{C_4}{C_2}\nu_z\;\left(\rho_z^2 - 2\rho_+^2 - 2\rho_-^2\right) + \frac{1}{16}\frac{C_3^2}{C_2^2}\nu_z\left(-15\rho_z^2 + 18\rho_+^2 + 18\rho_-^2\right),
    \label{eq:estatic}
\end{split}
\end{align}
where $\rho_z$ is the amplitude of the axial oscillation~\cite{Ketter2014}.
We use a 7-pole compensated Penning trap which has two pairs of correction electrodes to cancel the leading anharmonicities, see ref.~\cite{Heisse2019} for a similar design.
To this end we excite the radius of the magnetron motion $\rho_-$ and measure the dependencies $\nu_z(\rho_-^2, \rho_-^4)$ which is optimized to zero by adjusting the two voltages applied to the respective correction electrode pairs, see e.g. \ref{fig:tr_opt}.
Similarly, by measuring the dependence $\nu_z(\rho_z^2)$ for different correction electrode voltages, we can differentiate between the $C_3$ and $C_4$ terms in Eq.~(\ref{eq:estatic}).
The value of $B_2 \approx 1~\text{T}/\text{m}^2$ is measured by comparing its influence on $\nu_p$ and $\nu_z$.
All frequency shifts due to higher order field imperfections can be neglected at the current experimental precision.

The free cyclotron frequency $\nu_c$ is shifted due to the relativistic mass increase to smaller observed frequencies~\cite{Ketter2014}
\begin{equation}
    \nu_c^{\prime} = \frac{1}{\gamma}\nu_c \approx \left( 1 - \frac{1}{2}\frac{(2\pi\nu_c)^2\rho_+^2}{c^2} \right)\nu_c,
\end{equation}
where $c$ is the speed of light in vacuum and $\gamma$ is the Lorentz factor.
Typically we have small shifts $\gamma -1 < 10^{-11}$.
A hyperfine transition is on resonance, if the microwave frequency $\nu^{\prime}_{\text{MW}}$ in the ions rotating (rest) frame is equal to the transition frequency $\nu_i({B^{\prime}})$ at the ion's position.
In the rotating frame the magnetic field is perceived as the boosted quantity $B^{\prime} = \gamma B$ and the microwave frequency is shifted by the transverse Doppler effect to $\nu^{\prime}_{\text{MW}} = \gamma \nu_{\text{MW}}$.
If we use the observed (not corrected for relativistic shift) $\nu_c^{\prime}$ for our fit of $\langle\Delta^{\prime}_i\rangle$, the first order relativistic shift evaluates to a shift of the center value
\begin{equation}
    \langle\Delta^{\prime}_i\rangle = \langle\Delta_i\rangle - (\gamma - 1)\left(\nu_i - 2\nu_c\frac{\partial \nu_i}{\partial \nu_c}\right).
\end{equation}
Higher order effects, in particular the Thomas precession which contributes about $3\times 10^{-12}$ to $\Gamma_I$, can be neglected at our current precision.

\begin{figure}[!h]
\centering
    \includegraphics{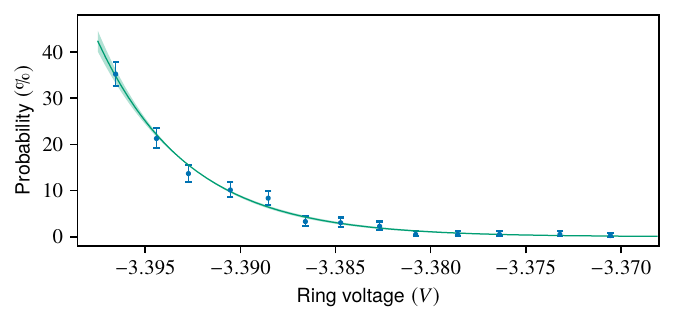}
    \caption{Measured distribution of voltages in the AT. The fit is performed via MLE. For details see text.}
    \label{fig:temp}
\end{figure}

The above shifts depend on the thermal radii of the ion, which are related to the temperature of the detection system at $\sim 4.2~\text{K}$.
In practice, the related temperature can be quite a bit higher, owing to e.g. heating from cryogenic amplifiers.
We determine the temperature of the axial detection system in the PT by measuring the energy distribution of the cyclotron mode in the AT.
During each cycle of our measurement we have to compensate for the strong $B_{2, \text{AT}} \approx 282~\text{kT}\text{m}^{-2}$ induced frequency shift, Eq.~(\ref{eq:B2nuz}), by adjusting the ring voltage to bring the ion into resonance with the AT detection system.
From an MLE fit to the distribution of ring voltages, see \ref{fig:temp}, we determine the temperature.
We use the mean temperature value determined from the three resonances and assign a conservative uncertainty of the largest difference between the measurements.
The value $T_z = 6.28(30)~\text{K}$ corresponds to the thermal amplitude $\rho_z = 35.41(85)~$\textmu m which is related to the modified cyclotron and magnetron amplitudes via $\rho_{\pm} = \sqrt{\nu_z/\nu_+}\rho_z$.

A shift of the motional frequencies arises from to the image charges induced on the trap surfaces due to the Coulomb force of the ion.
We calculate the relative shift on the cyclotron frequency to be $\delta\nu_c/\nu_c = 95.8(4.8)\times10^{-12}$, where the uncertainty of this shift is typically taken to be $5\%$~\cite{Schuh2019}.

The axial and radial frequencies are determined by fits of a thermal dip lineshape to the Fourier spectrum of our detection signal~\cite{WINELAND1975}.
This lineshape can be modelled with additional parameters, which account for e.g. amplification transfer functions, non-ideal noise short circuits by the ion, etc.
We estimate an upper bound for this systematic uncertainty on the cyclotron frequency to be $\sigma(\nu_c) = 10~\text{mHz}$.

Our detection system and all signal generators are connected to a rubidium atomic clock, which in turn is locked to a GPS reference clock.
Any relative frequency offset $\delta_{\text{ref}}$ of the clock leads to the following shift
\begin{equation}
    \Delta^{'}_i \approx \Delta_i + \delta_{\text{ref}}(\nu - \nu_c\frac{\partial \nu_i}{\partial \nu_c})
\end{equation}
The GPS locked rubidium clock, model FS725 from SRS, is rated at $10^{-12}$ accuracy and stability over days of measurement time, so a correlated and uncorrelated uncertainty $\sigma(\delta_{\text{ref}})_{\text{corr}} = \sigma(\delta_{\text{ref}})_{\text{uncorr}} = 10^{-12}$ among the three measured center values may exist and has to be taken into account.
While a common/correlated uncertainty of the reference only leads to an uncertainty of $\nu_{\text{HFS}}$, the uncorrelated uncertainty gives rise to shifts on all parameters, and especially, leads to large uncertainty for $\Gamma_I$.

In ref.~\cite{Moskovkin2006}, second-order corrections to the energy levels of the Zeeman and hyperfine splitting are discussed.
Including the non-zero nuclear electric quadrupole moment $Q(^9\text{Be}) = 0.0529(4)\times 10^{-28}\text{m}^{-2}$~\cite{Stone2021} leads to small shifts of the transition frequencies.
Comparing the frequencies as calculated from ref.~\cite{Moskovkin2006}, $\nu_{Q,i}$, with the classical Breit-Rabi derivation gives, $\nu_{Q, 1} - \nu_1 = -36.4(3)~\text{mHz}$, $\nu_{Q, 2} - \nu_2 = 5.69(4)~\text{mHz}$ and $\nu_{Q, 3} - \nu_3 = 4.36(3)~\text{mHz}$.

Table~\ref{tab:systematics} lists the statistical results and the systematic shifts.
\begin{table}[!h]
\small
  \caption{\label{tab:systematics}Statistical results and error budget. For more details, see text.}
    \begin{tabular}{l r r r}
    \hline
      & $\boldsymbol{\Gamma_e}$ & $\boldsymbol{\Gamma_I}$ & $\boldsymbol{\nu}_{\textbf{\text{HFS}}}$ \\\hline
      \rule{0pt}{2.5ex}\textbf{Statistical result} & $-5479.8633446(11)$ & $2.1354753839(11) \times 10^{-4}$ & $-12796971342.569(50)~\text{Hz}$ \\\hline
      \rule{0pt}{2.5ex}\textbf{Systematic shifts } & $/10^{-7}$  & $/10^{-14}$ & $/\text{mHz}$\\
      \rule{0pt}{2ex}Field imperfections & $-1(<1)$ & $0(<1)$ & $0(<1)$\\
      Relativistic & $0(<1)$ & $0(1)$ & $50(3)$\\
      Image charge & $-5(<1)$ & - & -\\
      Dip & $0(19)$ & - & -\\
      Frequency reference & $0(1)$ & $0(3)$ & $0(15)$\\
      Quadrupole moment & $-5(<1)$ & $-15(<1)$ & $11(<1)$\\
      Total shifts & $-11(19)$ & $-15(3)$ & $61(15)$\\
      \hline
      \rule{0pt}{2.5ex}\textbf{Corrected result} & $-5479.8633435(22)$ & $2.1354753854(11)\times 10^{-4}$ & $-12796971342.630(52)~\text{Hz}$\\\hline
    \end{tabular}
\end{table}
Additionally, we summarize the correlation via the Pearson correlation coefficients $\rho(X, Y) = \frac{\text{cov}(X, Y)}{\sigma(X)\sigma(Y)}$~\cite{Pearson1895}, see Table~\ref{tab:correlations}.
\begin{table}[!h]
\small
\centering
  \caption{\label{tab:correlations}Correlation coefficients of the fit parameters.}
    \begin{tabular}{l r r r}
    \hline
      \rule{0pt}{3ex} & statistical & systematic & combined\\ \hline
      \rule{0pt}{3ex}$\rho(\Gamma_e, \Gamma_I)$ & $0.48$ & $0.04$ & $0.24$\\
      \rule{0pt}{3ex}$\rho(\Gamma_e, \nu_{\text{HFS}})$ & $0.80$ & - & $0.38$\\
      \rule{0pt}{3ex}$\rho(\Gamma_I, \nu_{\text{HFS}})$ & $0.60$ & $0.14$ & $0.56$\\
      \hline
    \end{tabular}
\end{table}

\section{Supplement theory}

\subsection*{Hyperfine splitting and effective Zemach radius}
The energy difference between the two ground-state hyperfine sublevels of the $^9$Be$^{3+}$ ion at zero magnetic field can be parametrized by
\begin{align}
E_{\mathrm{HFS,theo}} = 2 h \nu_{\mathrm{HFS}} =  E_{\text{F}} (  A(Z\alpha) +  \delta_{\mathrm{recoil,pt}} + \delta_{\mathrm{QED,pt}} +  \delta_{\mu \mathrm{VP}} + \delta_{\mathrm{had\,VP}} + \delta_{\mathrm{FS}}) \, .
\label{eq:HFS1s}
\end{align}
For the $1s$ state and a nuclear spin of $I=3/2$~\cite{Stone2019}, the non-relativistic Fermi energy $E_{\text{F}}$ is given by~\cite{Beier00}
\begin{align}
E_{\text{F}} = \alpha g_I \frac{m_e}{m_p} \frac{8}{3} m_e c^2 (Z\alpha)^3 \mathcal{M} \, .
\end{align}
Here, $\alpha$ is the fine-structure constant.

The relativistic factor can be calculated analytically, using wave functions for the model of a point-like nucleus. 
For the $1s$ state, it is~\cite{Beier00}
\begin{align}
A(Z\alpha) = \frac{1}{\gamma (2 \gamma -1)} \, ,
\end{align}
with $\gamma = \sqrt{1 - (Z\alpha)^2}$.

The finite nuclear mass (recoil) correction is partially taken into account by the mass factor $\mathcal{M} = \left( 1 + m_e/M_N \right)^{-3}$~\cite{Beier00}, with the nuclear mass $M_N$. 
For recoil corrections beyond the mass factor and to leading order in $Z\alpha$, we use the formula developed for systems with point-like nuclei (e.g. muonium)~\cite{Karshenboim1996}. 
Higher-order (in $Z\alpha$) recoil corrections are treated as in refs.~\cite{Bodwin1985, Bodwin1988, Pachucki2023_2}.
Since, to our knowledge, radiative recoil corrections to HFS have not been calculated for nuclei with spin different from $1/2$, we used the point-nucleus formula for spin-$1/2$ from ref.~\cite{Karshenboim1996} as the uncertainty due to uncalculated radiative recoil corrections.

The QED corrections can be parametrized as~\cite{Kinoshita1998}
\begin{align}
\delta_{\mathrm{QED}} = a_e + \delta_{\mathrm{QED,binding}} \, ,
\end{align}
with the free electron anomaly $a_e$~\cite{Codata2018}. 
One-loop binding corrections were calculated to all orders in $Z\alpha$. 
Specifically, the one-loop self-energy was calculated as the sum of the perturbative $Z\alpha$ expansion formula as given in refs.~\cite{Beier00,Yerokhin2008} and the tabulated higher-order term from ref.~\cite{Yerokhin2008}.
In ref.~\cite{Heiland2023}, we calculated the vacuum polarization (VP) correction for the model of a point-like nucleus, with the VP loop in the Uehling approximation~\cite{QFT}. 
The electric-loop contribution was calculated using the electron wave function perturbed by the nuclear magnetic field, as given in ref.~\cite{ShabaevVirial,Karshenboim2000VPHFS}. 
The magnetic loop correction was calculated following refs.~\cite{Beier00,Schneider1994}.
We found our VP results for various nuclear charge numbers to be in excellent agreement with results from ref.~\cite{Beier00,Karshenboim2000VPHFS}. 
The above SE and VP results were obtained for the point-like nuclear model.

Two-loop binding corrections have been calculated up to order $(Z\alpha)^2$~\cite{Volotka2006,Kinoshita1998,Karshenboim2002}. 
Apart from the (for the purpose of our calculation negligibly small) uncertainty of $a_e$~\cite{Codata2018}, we took into account the uncertainty of the one-loop self-energy correction as given in ref.~\cite{Yerokhin2008} and the given uncertainties of the two-loop $Z\alpha$ expansion coefficients~\cite{Karshenboim2002,Kinoshita1998}. 
We estimated the uncertainty of the QED parameter due to uncalculated two-loop binding corrections of orders $(Z\alpha)^3$ and higher as $2 \left( \alpha/\pi \right)^2 (Z\alpha)^3 \log \left[ (Z\alpha)^{-2} \right] \approx 1.9 \times 10^{-9}$. 
In a similar way, the uncertainty due to uncalculated 3-loop binding corrections was estimated as $2 \left( \alpha/\pi \right)^3 (Z\alpha)  \approx 7.3 \times 10^{-10}$.
In total, we obtain the QED correction parameter $\delta_{\mathrm{QED}} = 0.000 \, 722 \, 219(12)$.

The parameters for the muonic and hadronic vacuum polarization corrections $\delta_{\mu \mathrm{VP}}$ and $\delta_{had \mathrm{VP}}$ in Table~\ref{tab:HFS} correspond to the sum of the muonic/hadronic Uehling potential correction and the leading-order magnetic loop correction for the point-like nuclear model and are taken from ref.~\cite{Heiland2023}. 
We include both parameters with a relative uncertainty of 10\% to account for uncalculated light-by-light scattering type muonic and hadronic vacuum polarization contributions.

All contributions discussed above assume a point-like treatment of the nucleus and are summarized in Table~\ref{tab:HFS}.
\begin{table}
\small
\centering
\caption{\label{tab:HFS} Contributions to the ground-state hyperfine splitting in $^{9}$Be$^{3+}$. All contributions listed here assume a point-like nucleus.}
\begin{tabular}{lrll}
\hline
Contribution& &  &Following refs. \\ 
\hline 
Fermi energy& $E_{\text{F}}/h = $ & -25 558 815.5~kHz& \cite{Codata2018}\\\hline
Relativistic&$A(Z\alpha)-1=$ & \phantom{-}0.001 279 577&\cite{Shabaev1994,Beier00} \\ 
QED &$\delta_{\mathrm{QED}}=$ & \phantom{-}0.000 722 219(12)& \cite{Codata2018,Yerokhin2008,Beier00,Kinoshita1998,Karshenboim1997} \\ 
Recoil&   $\delta_\mathrm{recoil}=$ & -0.000 017 38(24)  &      \cite{Karshenboim1996,Bodwin1985,Bodwin1988} \\ 
Muonic VP& $\delta_{\mu \mathrm{VP}}=$ & \phantom{-}0.000 000 78(8) & \cite{Heiland2023} \\
Hadronic VP &$\delta_{had \mathrm{VP}}=$ & \phantom{-}0.000 001 11(11)& \cite{Heiland2023} \\
\hline
\end{tabular} 
\end{table}
We incorporate all corrections to the HFS related to the finite size of the nucleus (nuclear structure) into the effective Zemach radius~$\tilde{r}_{\mathrm{Z}}$, defined following ref.~\cite{Puchalski2014} as,
\begin{align}
\delta_{\mathrm{FS}} = - 2Z\tilde{r}_Z/a_0 \, .
\end{align}
Apart from the finite size of the nuclear charge distribution, this includes the dominant Bohr-Weisskopf effect, nuclear structure dependent recoil effects, and additional QED finite-size cross-terms as well as nuclear polarization corrections.
From the experimental value of the hyperfine splitting in our work, assuming $E_{\mathrm{HFS,theo}} = E_{\mathrm{HFS,exp}}$, we find the effective Zemach radius to be $\tilde{r}_Z=4.048(2)$ fm. 
The uncertainty is dominated by uncalculated radiative recoil contributions (see discussion above).

\subsection*{Weighted difference of the hyperfine splittings}

For $^9$Be, with a new value for the HFS in the hydrogen-like charge state in our work, and a high-precision measurement of the HFS in the lithium-like charge state in ref.~\cite{Shiga2011}, we put forward a weighted difference of hyperfine splittings in such a way that nuclear effects are strongly suppressed, see ref.~\cite{Shabaev2001HFS}, and compare the experimental value for the weighted difference with the theoretical value for the weighted difference. 

The theoretical value for the hyperfine splitting of the Li-like Be ion can be parametrized following ref.~\cite{Puchalski2014}
\begin{align}
\begin{split}
E_{\text{ HFS,2s}} = & E_{\text{F,2s}} (  A_{2s}(Z\alpha)  - 2 \alpha Z  \tilde{r}_{\mathrm{Z}} +  \delta_{\mathrm{recoil,pt}} + \delta_{\mathrm{QED,2s,pt}}) = \\
& \frac{m_e^2 g_I}{Z m_p} \left( \frac{g_e}{2} \alpha^4 A^{(4)} + \sum \limits_{n=5}^\infty \alpha^n A^{(n)} \right) \, , 
\label{eq:HFS2s}
\end{split}
\end{align}
with the free electron g-factor $g_e = 2 (1 + a_e)$.
The first parametrization was chosen to resemble our parametrization of the HFS in the hydrogen-like charge state from the previous section. 
The second parametrization is taken from ref.~\cite{Puchalski2014}. 
In that work, explicit values are given for the coefficients $A^{(4)}$ to $A^{(7)}$, with $A^{(7)}$ being an estimate based on one-electron bound QED. 
We first discuss how coefficients from both parametrizations are related to each other.

The leading-order contribution in the non-relativistic limit is given by
\begin{align}
E_{\text{F,2s}} = \frac{m_e^2 g_I}{Z m_p} \alpha^4 A^{(4)} \, ,
\end{align}
with the numerical value for $A^{(4)}$ tabulated in ref.~\cite{Puchalski2014}. 
Relativistic corrections to the leading order are parametrized as
\begin{align}
A_{2s}(Z\alpha) = 1 + \alpha^2 \frac{A^{(6)}_{\text{rel}}}{A^{(4)}} \, .
\end{align}
Here, $A^{(6)}_{\text{rel}}$ corresponds to a part of the total term $A^{(6)}$ which contains relativistic and QED terms as follows,
\begin{align}
\begin{split}
A^{(6)} & = A^{(6)}_{\text{rel}} + A^{(6)}_{\text{R}}, \\
A^{(6)}_{\text{rel}} & = A^{(6)}_{\text{AN}} + A^{(6)}_{\text{B}} + A^{(6)}_{\text{C}}. 
\end{split}
\end{align}
In ref.~\cite{Puchalski2014}, numerical values for the relativistic corrections $A^{(6)}_{\text{AN}}, A^{(6)}_{\text{B}}$ and $A^{(6)}_{\text{ C}}$ as well as the QED term $A^{(6)}_{\text{R}}$ are tabulated. 
The term $A^{(6)}_{\text{R}}$ corresponds to one-loop QED corrections of $\mathcal{O} (Z\alpha)$. 
The term $A^{(7)}$ corresponds to the sum of one-loop QED terms of $\mathcal{O} \left( (Z\alpha)^2 \right)$ and two-loop QED terms of $\mathcal{O} (Z\alpha)$. 
With this, we find the following QED parameter
\begin{align}
\delta_{\mathrm{QED,2s,pt}} = a_e + \alpha^2 \frac{A^{(6)}_{\text{ R}}}{A^{(4)}} + \alpha^3 \frac{A^{(7)}}{A^{(4)}} \, .
\end{align}

Finally, the parameter $A^{(5)}$ corresponds to the sum of nuclear finite size and recoil corrections as follows
\begin{align}
\alpha \frac{A^{(5)}}{A^{(4)}} =  -2 m_e Z \alpha \tilde{r}_Z + \delta_{\mathrm{recoil,pt}} \, .
\end{align}
The recoil correction parameter for point nuclei is identical for the hydrogen-like and lithium-like systems.

In order to combine the hydrogen-like and the lithium-like hyperfine splitting values in a way to efficiently cancel nuclear effects, we may use the following weighted difference
\begin{align}
E_{\text{HFS}}^\xi = E_{\text{HFS,2s}} - \xi E_{\text{HFS,1s}} \, ,
\end{align}
with the weight factor
\begin{align}
\xi = \frac{E_{\text{F,2s}}}{E_{\text{F}}} = \frac{3 A^{(4)}}{8 Z^4 \mathcal{M}} = 0.048 \, 818 \, 910 \, 46 \, .
\end{align}
With this weight factor, the effective Zemach radius corrections from lithium-like and hydrogen-like hyperfine splitting cancel exactly.

As a \textit{side effect} of the choice of weight factor, the leading contributions, represented by $E_{\text{F}}$ and $E_{\text{F,2s}}$ in the H-like and Li-like charge state respectively, cancel exactly.
With this, the weighted difference is several orders of magnitude smaller than the individual HFS values.
Furthermore, leading QED corrections due to the electron's magnetic moment anomaly $a_e$ and the one-loop QED term of order $Z\alpha$ cancel in the weighted difference. 
With this, the contributions that do remain in the weighted difference are relativistic corrections according to $A^{(6)}_{\text{rel}}$ in the lithiumlike hyperfine splitting and its corresponding hydrogenlike term, as well as higher-order QED terms $A^{(7)}$.  

The leading recoil corrections obtained for the point nucleus model cancel exactly.
Moreover, recent recoil calculations~\cite{Pachucki2022} imply that the leading-order part of the recoil parameters in equations \eqref{eq:HFS1s} and \eqref{eq:HFS2s} is identical even for extended nucleus calculations. 
We therefore assume an exact cancellation of the leading recoil contribution in the weighted difference.

The uncertainty due to uncalculated higher-order recoil corrections in the weighted difference was estimated based on the higher-order contribution to the recoil correction to the specific difference $D_{21}$ of hyperfine splittings of the $2s$ and $1s$ states, as given in ref.~\cite{Karshenboim2002}.

For the uncertainty due to uncalculated higher-order nuclear size corrections, we estimated the total nuclear size charge and magnetization distribution corrections for both $1s$ and $2s$ states using semi-analytic wavefunctions for the bound electron~\cite{Patoary2018} and formulas from refs.~\cite{Shabaev1994,Volotka2005}. 
We estimate an uncertainty due to higher-order finite size corrections to the weighted difference as $\frac{m_e^2 g_I}{2 Z m_p} A^{(4)} \vert \delta_{FS,2s} - \delta_{FS,1s} \vert$.

Considering that the lithiumlike theory is developed up to order $m \alpha^7$, in the following, we estimate uncertainties of the weighted difference that may arise due to uncalculated higher-order QED terms. 
Explicit calculations of the one-loop (electronic) vacuum polarization correction for the $1s$ and 2s states (the latter in the one-electron approximation) for both extended and point-like nuclei imply very similar VP-FS cross terms in both cases~\cite{Heiland2023}. 
With this, we estimate the uncertainty of the weighted difference due to QED-FS cross terms as $2 \frac{m_e^2 g_I}{2 Z m_p} A^{(4)} \vert \delta_{\mathrm{VP,FS,}2s} - \delta_{\mathrm{VP,FS,}1s} \vert$.

All estimated uncertainties due to higher-order finite size or recoil effects are at least one order of magnitude smaller than the estimated uncertainty of the Li-like QED theory as given in ref.~\cite{Puchalski2014}. 
For the total theoretical value of the weighted difference, taking into account the hydrogenlike QED theory up to order $m \alpha^7$, we find
\begin{align}
E_{\text{HFS,theo}}^\xi/h = -542.7(7.2)\, \mathrm{kHz} \, .
\end{align}
The experimental value for the weighted difference is determined by inserting the experimental HFS values for the H-like ion from this work, and for the Li-like system from ref.~\cite{Shiga2011}.

\begin{table}
\small
\centering
\begin{tabular}{l r rl rl }
\hline 
Order & $\Delta E_{1s}$ & $\Delta E_{2s}$ && $E_{\text{HFS}}^\xi$ & \\ 
\hline 
$m \alpha^4$, non-rel & -25 558 815.5 & -1 247 753.5 && 0.0 & \\ 
$m \alpha^4$, $a_e$ & -29 639.3 & -1 447.0 && 0.0 & \\  
$m \alpha^5$, Recoil & 421.0 & 20.6 && 0.0 & \\ 
$m \alpha^6$, rel & -32 665.0 & -2 122.3 && -527.6 & \\ 
$m \alpha^6$, 1-loop QED & 9836.8 & 480.3 && 0.0 & \\ 
$m \alpha^7$, 1-loop QED & 1211.3 & 44.0 &(7.2) & -15.1 & (7.2) \\ 
$m \alpha^7$, 2-loop QED & -9.8 & -0.5 && 0.0 & \\ 
\hline 
\end{tabular}
\caption{Contributions of different orders in $\alpha$ to the hyperfine splitting in H-like and Li-like $^9$Be ions and to the weighted difference, in kHz.}
\end{table}

\subsection*{Diamagnetic shielding}
The combined effect of the external and the nuclear magnetic fields leads to a splitting of the $1s$ state into multiple magnetic sublevels. 
To first order in perturbation theory, these sublevels are described with the standard Breit-Rabi formulas~\cite{FeynmanLectures}.
In ref.~\cite{Moskovkin2006}, several second-order corrections to the Breit-Rabi formulas were calculated. 
The second-order Zeeman shift turns out to be identical for all sublevels and therefore does not have an impact on frequencies of transitions between sublevels. 
We also estimated the third order Zeeman splitting by extrapolating tabulated results from ref.~\cite{Varentsova2017} to low $Z$.
Even for a magnetic field of 7~T, the third-order Zeeman shift corresponds to a correction to the bound-electron's $g$-factor of $4 \times 10^{-15}$ and is therefore negligibly small. 

Apart from that, the modification of the Breit-Rabi formulas due to nuclear magnetic dipole shielding as well as the nuclear electric quadrupole shielding were calculated in ref.~\cite{Moskovkin2006}. 
The magnetic dipole shielding corresponds to Feynman diagrams with two magnetic interactions, namely one external magnetic field and one nuclear magnetic field line \cite{Yerokhin2012}.
Assuming the model of a point-like nucleus, the leading value of the shielding constant can be given analytically as~\cite{Grotch1971,Close1971,Faustov1970,Moore1999,Pyper1999,Pyper1999II}.
\begin{align}
\sigma^{(0)} = -\frac{4 \alpha Z \alpha}{9} \left( \frac{1}{3} - \frac{1}{6(1+\gamma)} + \frac{2}{\gamma} - \frac{3}{2\gamma-1} \right) \, .
\end{align}

The combined finite size and Bohr-Weisskopf correction to the shielding constant were calculated using formulas from ref.~\cite{Pachucki2023_3} (Results from this formula were found to be consistent with numerical results tabulated for low-Z ions in ref.~\cite{Yerokhin2012}).

Recoil corrections to the shielding constant were calculated using formulas from refs.~\cite{Eides1997,Pachucki08,Yerokhin2012,Pachucki2023}
\begin{align}
\begin{split}
\sigma_{\mathrm{rec}} & = - \frac{\alpha Z \alpha}{3} \frac{m_e}{M_N} \left( 1 + \frac{g_N - 1}{g_N} \right) \, ,  \\
g_N & = \frac{M}{Z m_p} g_I \, .
\end{split}
\end{align}

The QED correction to the shielding constant was calculated to lowest order in $Z\alpha$ in ref.~\cite{Wehrli2021}. 
We estimate the QED contribution as given in Table~\ref{tab:shielding} to all orders in $Z\alpha$, by extrapolating higher-order terms from ref.~\cite{Yerokhin2012} to low $Z$. 
The uncertainty given in Table~\ref{tab:shielding} takes into account the difference between our extrapolated all-order result and the $Z\alpha$ expansion result~\cite{Wehrli2021}, as well as an estimation of uncalculated VP contributions~\cite{Yerokhin2012}. 

\begin{table}
\small
\centering
\caption{\label{tab:shielding}Contributions to the shielding constant in $^{9}$Be$^{3+}$.}
\begin{tabular}{llllllll}
\hline
Contribution &  & $\sigma(^{9}$Be$^{3+}$) &  Following refs. \\ 
\hline 
Leading value &  all-order $Z\alpha$ & \phantom{-}0.000 071 165 01  &  \cite{Grotch1971,Close1971,Faustov1970,Moore1999,Pyper1999,Pyper1999II} \\ 
Finite size& &  -0.000 000 000 16(3) & \cite{Pachucki2023_3} \\ 
One-loop QED & all-order $(Z \alpha)$&  \phantom{-}0.000 000 000 23(13) & \cite{Wehrli2021,Yerokhin2012} \\ 
Recoil &  $\mathcal{O } \left( Z \alpha^2 \frac{m}{M} \right)$   & -0.000 000 011 11(1)  &  \cite{Grotch1971,Close1971,Faustov1970,Yerokhin2012} \\ 
\hline
Sum &   &  \phantom{-}0.000 071 153 97(14) & \\
\hline
\end{tabular} 
\end{table}

\subsection*{Bound-electron $g$-factor}
The theory of the bound-electron $g$-factor is very similar to the case of the $^3$He$^+$ ion \cite{Schneider2022}. 
The leading zero-loop $g$-factor contribution was first calculated for bound electrons in 1928~\cite{Breit1928}. 
The contribution of QED Feynman diagrams with closed loops can be parametrized as the sum of the free electron's anomaly~\cite{Codata2018} and so-called binding corrections. 
For the light Be ion, binding corrections calculated in the framework of a perturbative expansion in the electron-nucleus interaction, with the expansion parameter being $Z\alpha$, are found to converge well. 
For the two-loop correction, already terms of order $(Z\alpha)^4$ and higher~\cite{Pachucki05,Czarnecki16,Czarnecki2017,Czarnecki2020,Yerokhin2Loop2013} are found to be smaller than $10^{-10}$, as well as three-loop binding corrections of order $(Z\alpha)^2$. 
The higher-order contributions to the one-loop self-energy correction is based on the recent high-precision evaluation of these Feynman diagrams to all orders in $Z\alpha$~\cite{Yerokhin2017}. 
The Wichmann-Kroll contribution to the electric loop vacuum polarization correction~\cite{Beier00} as well as the magnetic loop vacuum polarization correction~\cite{Lee2005} turn out to be too small to contribute at the given level of precision.

Unlike the case of the $^3$He$^+$ ion, the biggest uncertainty of theory contributions to the $g_s$- factor in $^9$Be$^{3+}$ originates from uncalculated higher-order QED corrections at the two-loop level~\cite{Czarnecki2017}.
It is slightly larger than the uncertainty of the finite nuclear size correction, which was calculated nuclear model independently using formulas and tabulated parameters from refs.~\cite{Yerokhin2013,Zatorski12}. 
The uncertainty of the finite size contribution originates from the uncertainty of the nuclear root mean square radius as specified in ref.~\cite{Angeli2013}.
This means that a $g$-factor experiment with sufficient accuracy, combined with an independent mass measurement of the Be$^{3+}$ ion, would allow for the direct and independent determination of the $r_{\mathrm{rms}}$ radius with an uncertainty which is comparable to the best available value~\cite{Angeli2013}.

\begin{table}
\small
\centering
\caption{\label{tab:gkompakt}Contributions to the bound-electron $g$-factor in $^{9}$Be$^{3+}$. All digits given are significant.}
\begin{tabular}{rlllllll}
\hline
Contribution & & $^{9}$Be$^{3+}$ &  Following refs. \\ 
\hline 
Dirac value       &                                         & \phantom{-}1.999 431 864 5  &  \cite{Breit1928,Beier00} \\ 
Finite size       &                                         & \phantom{-}0.000 000 000 1  & \cite{Yerokhin2013,Angeli2013,Trouyet2019,Moutet2020,GRASP2018,Zatorski12,Karshenboim2018} \\ 
1-loop QED        & $(Z \alpha)^0$                          & \phantom{-}0.002 322 819 5  & \cite{Schwinger1948,Codata2018} \\ 
                  & $(Z \alpha)^2$                          & \phantom{-}0.000 000 329 8  &  \cite{Grotch1971,Close1971,Faustov1970,Czarnecki00} \\ 
                  & $(Z \alpha)^4$                          & \phantom{-}0.000 000 023 3  &  \cite{Pachucki2004oneloop,Pachucki05} \\ 
                  & $(Z \alpha)^{5+}$ SE                    & \phantom{-}0.000 000 001 1  &  \cite{Yerokhin04,Beier00} \\ 
                  & $(Z \alpha)^{5+}$ VP-EL, Uehling       & \phantom{-}0.000 000 000 1  & \cite{Karshenboim2001,QFT,Beier00,Karshenboim2005} \\ 
2-loop QED        & $(Z \alpha)^0$                          & -0.000 003 544 6   &  \cite{Petermann1957,Sommerfield1958,Codata2018} \\ 
                  & $(Z \alpha)^2$                         & -0.000 000 000 5    & \cite{Grotch1971,Close1971,Faustov1970,Czarnecki00} \\ 
$\geq$ 3-loop QED & $(Z \alpha)^0$                          & \phantom{-}0.000 000 029 5      & \cite{Codata2018,Laporta1996,Kinoshita2007,Kinoshita2012,Laporta2017,Kinoshita2017} \\ 
Recoil            & $\frac{m}{M}$, all-order $Z \alpha$    & \phantom{-}0.000 000 051 9     &  \cite{Shabaev02,Beier00,AME2020,GRASP2018,NIST} \\ 
\hline
Sum               &                                        & \phantom{-}2.001 751 574 7  & \\
\hline
\end{tabular} 

\end{table}

\section{Comparison plot data}
The plotted data from Figure~4 of the main text is summarized in Table~\ref{tab:comp_data}.

For the HFS of the hydrogen isotopes, we use the theoretical QED calculation from ref.~\cite{Pachucki2023_2} to evaluate the point-like contribution.
The $g$-factor values, the fine-structure constant, which significantly contributes to the relative uncertainties of the zero-field splitting by about $3\times10^{-10}$, and the Rydberg constant are taken from CODATA~\cite{Codata2018}.
Compared to our calculations for $\HBe$, no contributions and associated uncertainties due to muonic vacuum polarization, hadronic vacuum polarization, and leading order recoil were included for H, D and T.
Thus, for a better comparison with our results for $\HBe$, the same calculation is applied to $\HBe$ as well, see the Table.

For tritium, the experimental result of the ratio $\nu_{\text{HFS}}(\text{T})/\nu_{\text{HFS}}(\text{H})$, ref.~\cite{Mathur1967} is combined with the hydrogen result, ref.~\cite{Hellwig1970}.

For $^{209}\text{Bi}^{82+}$ the theoretical result from ref.~\cite{Shabaev2000_Bi} is rescaled with the new value of its nuclear magnetic moment, ref.~\cite{Skripnikov2018}.
\begin{table}[!h]
  \scriptsize
  \caption{\label{tab:comp_data}Data for Fig.~4. In the theoretical contribution of the hyperfine splittings, the contribution to the uncertainty of the experimental value of $g_I$ is given if it is significant. The same applies to $\alpha$ and $m_e$. If no discrimination is made for the nuclear contribution, only the theoretical uncertainty is significant. The second line for $\HBe$ uses theory calculations from ref.~\cite{Pachucki2023_2}, which are used for the hydrogen isotopes H, D, T as well. For $\HBe$ these do not include the same contributions as our calculations (thus the deviation of the values), and are only meant for the comparison with the hydrogen isotopes. For details, see text.}
    \begin{tabular}{l r r r}
    \hline
    System & Exp. value & Theo. point nucl. value & Nucl. contrb.\\
    \hline\hline
    HFS & & & \\\hline
    $\HBe$ $1s$ & $-12796971.342629(52)$~kHz~[tw] & $-12804791.6(3.5)$~kHz~[tw] & $7820.2(3.5)$~kHz\\
     &  & $-12804973.08(20)$~kHz~\cite{Pachucki2023_2} & $8001.74(20)$~kHz\\
    H $1s$ & $1420405.751768(2)$~kHz~\cite{Hellwig1970} & $1420452.3711(14)(4)_{g_I}(4)_{\alpha}$~kHz~\cite{Pachucki2023_2} & $-46.6193(15)$~kHz\\
    D $1s$ & $327384.3525222(17)$~kHz~\cite{Wineland1972} & $327339.21932(33)(84)_{g_I}(10)_{\alpha}$~kHz~\cite{Pachucki2023_2} & $45.13321(91)$~kHz\\
    T $1s$ & $1516701.4707745(74)$~kHz~\cite{Mathur1967, Hellwig1970} & $1516760.1094(15)(31)_{g_I}(5)_{\alpha}$~kHz~\cite{Pachucki2023_2} & $-58.6386(34)$~kHz\\
    $^3\text{He}^+$ $1s$ & $-8665649.86577(26)$~kHz~\cite{Schneider2022} & $-8667379.5(1.3)$~kHz~\cite{Schneider2022} & $1729.7(1.3)$~kHz\\
    $^{209}\text{Bi}^{82+}$ $1s$ & $5085.03(1)$~meV~\cite{Ullmann2017} & $5137(3)_{g_I}$~meV~\cite{Shabaev2000_Bi} & $-52(3)$~meV\\
    $\LBe$ $2s$ & $-625008.837044(12)$~kHz~\cite{Shiga2011} & $-625389.2(3.6)$~kHz~\cite{Puchalski2014} & $380.4(3.6)$~kHz\\
    $^6\text{Li}^+$ $2s$ & $3001805.1(5.1)$~kHz~\cite{Sun2023} & $3002617.5(1.5)(1.4)_{g_I}$~kHz~\cite{Pachucki2023_2} & $-812.4(2.0)_{\text{theo}}(5.1)_{\text{exp}}$~kHz\\
    $^7\text{Li}^+$ $2s$ & $7926990(17)$~kHz~\cite{Guan2020} & $7929977.4(3.9)(1.3)_{g_I}$~kHz~\cite{Pachucki2023_2} & $-2987(4)_{\text{theo}}(17)_{\text{exp}}$~kHz\\ \hline
    other & & & \\\hline 
    \textmu H Lamb shift & $202.3706(23)$~meV~\cite{Antognini2013} & $206.0336(15)$~meV~\cite{Antognini2013} & $-3.663(15)_{\text{theo}}(23)_{\text{exp}}$~meV\\
    \textmu D Lamb shift & $202.8785(34)$~meV~\cite{Pohl2016} & $228.7766(10)$~meV~\cite{Pohl2016} & $-25.8981(10)_{\text{theo}}(34)_{\text{exp}}$~meV\\
    $g_s(^{118}\text{Sn}^{49+})$ & $1.910562059(9)$~\cite{Morgner2023} & $1.91054733(30)$~\cite{Morgner2023} & $14.73(30)\times 10^{-6}$ \\
    \hline
    \end{tabular}
\end{table}

\printbibliography